\documentclass[12pt,twocolumn]{article}
\usepackage{epsfig}

\title{Ordering temperatures and critical exponents in Ising spin glasses}

\author{L.W. Bernardi, S. Prakash and I.A. Campbell\\
        Physique des Solides,\\
        Universit\'e Paris Sud,\\
        91405 Orsay, France}

\begin{document}

\onecolumn
\maketitle

\begin{abstract}
We propose a numerical criterion which can be used to obtain accurate and
reliable values of the ordering temperatures and critical exponents of spin
glasses. Using this method we find a value of the ordering temperature for
the $\pm J$ Ising spin glass in three dimensions which is definitely non-zero
and in good agreement with previous estimates. We show that the critical
exponents of  three dimensional Ising spin glasses do not appear to obey 
the usual universality rules.
\end{abstract}


        The full explanation of the universality rules for critical
exponents in second order transitions through the renormalization group
theory is one of the most impressive achievements of statistical physics.
The universality rules for such transitions state that the critical
exponents depend only on the space dimension $d$ and a few basic parameters :
the number of order parameter components $n$, the symmetry and the range of
the Hamiltonian \cite{1}. No other parameters are pertinent. In fact it is 
known that there are exceptions to universality - in certain two dimensional 
($2d$) Ising systems with regular frustration the critical exponents vary
continuously with the value of a control parameter \cite{2}. As far as we are
aware, no results of this type have been reported in any three dimensional
($3d$) family of Ising systems; it has been tacitly assumed that
non-universality is very exceptional.

        As compared to standard second order transitions, the situation
concerning Ising Spin Glasses (ISGs) is much less clear;  indeed the
history of ISG simulations has been plagued by technical difficulties
associated with long relaxation times. For two decades the very existence
of a finite temperature transition in the $3d$ ISG has been hotly contested;
as it is obviously essential to have a reliable value of the ordering
temperature before obtaining accurate critical exponent estimates, it has
been difficult to make stringent numerical tests of universality in $3d$
ISGs.

        We will present a numerical criterion which can in favourable cases
provide precise and reliable values for the ordering temperature $T_g$ and for
the critical exponents of a spin glass, with a moderate level of
computational effort. If an independent estimate of the ordering
temperature is available the criterion leads to a convenient method for
estimating the exponents. We study $3d$ ISGs with various sets of
interactions and we conclude from the data that  the $3d$ 
$\pm J$ interaction ISG has a well defined non-zero Tg which can 
be estimated accurately, and that universality in the usual sense does 
not hold in $3d$ ISGs.

        It would appear probable that glassy transitions in general have a
much richer critical behaviour than have conventional second order
transitions.

        Thus, technically the most difficult problem in numerical ISG
studies is the correct identification of the transition temperature $T_g$. For
the $3d$ ISG with  random $\pm J$ near neighbour interactions on a simple cubic
lattice, which has been the subject of a considerable ammount of work, $T_g$
has been estimated in two ways. Ogielski \cite{3} studied in massive 
simulations the divergence of the spin glass susceptibility, of the 
correlation length, and of the relaxation time of the autocorrelation function
\begin{equation}
                q(t)=\left[<S_i(t)S_i(0)>\right]      
\label{eq:1}
\end{equation}
in order to estimate $T_g$ and the critical exponents. However his analysis
has been questioned because of the possibility of ambiguities in the manner
of identifying a divergence, if non-conventional temperature dependencies
are invoked \cite{4}. Bhatt and Young \cite{5} used a finite size scaling 
technique; they measured the Binder cumulant for the fluctuations of the 
equilibrium autocorrelation function
\begin{equation}
                gL=\frac{1}{2}\left[ 3 - \frac{< q^4>}{<q^2>^2}\right]
\label{eq:2}
\end{equation}
as a function of sample size $L$. The curves $g_L(T)$ for different $L$ should
all intersect at $T_g$; in the $3d$ $\pm J$ ISG case the curves indeed 
intersected but did not appear to fan out below the apparent $T_g$. Only 
recently have intensive numerical studies shown that a weak fanning out at low
temperatures really does occur \cite{6,7}. Even with results of high 
statistical accuracy to hand, Kawashima and Young \cite{6} give a number of 
caveats concerning the interpretation of their own data.

        We will describe an alternative criterion for determining $T_g$.
First, scaling rules tell us \cite{3} that for a large sample in thermal
equilibrium at $T_g$ the relaxation of the autocorrelation function takes the
form
\begin{equation}
                q(t)=\lambda t^{-x}
\label{eq:3}
\end{equation}

with the exponent $x$ related to the standard static and dynamic exponents 
$\eta$ and $z$ through
\begin{equation}
                x=\frac{(d-2+\eta)}{2z}.
\label{eq:4}
\end{equation}

        Secondly, the out of equilibrium relaxation of two randomly chosen
replicas $A$ and $B$ of the same sample towards equilibrium at $T_g$ depends on
another combination of the same exponents \cite{8}. The out of equilibrium spin
glass susceptibility is defined as
\begin{equation}
                \chi'_{SG}(t)=\left[<S_i^A(t)S_i^B(t)>^2\right]
\label{eq:5}
\end{equation}
and it increases with time as 
\begin{equation}
                t^h \ \mathrm{with}\  h = \frac{2-\eta}{z}.
\label{eq:6}
\end{equation} 
        Suppose we take $\{T_i\}$, a series of trial values for $T_g$; from
measurements of $x$ and $h$ on large samples at each $T_i$ we can deduce from
equations \ref{eq:4} and \ref{eq:6} a set of apparent or effective exponents
\begin{equation}
        \eta_1(T)=\frac{4x - h(d-2)}{2x+h}
\label{eq:7}
\end{equation} 
\begin{equation}
         z(T)=\frac{d}{2x+h}.                                         
\label{eq:8}
\end{equation} 

        Finally in another set of simulations on the same system at
different [small] sample sizes $L$, from standard finite size scaling rules
\cite{5} for the fluctuations in the autocorrelation function in equilibrium at
$T_g$ we have
\begin{equation}
                L^{d-2}<q^2>\   \propto\   L^{-\eta}
\label{eq:9}
\end{equation}
        If we again take a series of trial values of $T_g$ and fit the results
using this form at each $T_i$ we will obtain a second series of apparent
exponent values $\eta_2(T)$. (This type of fit will only be appropriate 
close to and below $T_g$; at higher $T$ another factor appears on the 
right hand side \cite{5}).

        We now plot $\eta_1(T)$ and $\eta_2(T)$ against $T$; for consistency 
the curves must intersect at the point [$\eta$, $T_g$] which represents the 
true critical exponent $\eta$ and ordering temperature $T_g$ of the system.
 At this temperature and this temperature only the functional forms of 
equations \ref{eq:3}, \ref{eq:6} and \ref{eq:9} should be exact; at 
neighbouring temperatures  these forms are only approximate but close to 
$T_g$ they will be adequate to parametrise the numerical data. Once $T_g$ 
is fixed by the intersection  we can obtain $z$ using the $z(T)$ curve given 
above, and with known  $\eta$ and $T_g$ we can go on to fit $<q^2>$ data for 
temperatures above $T_g$ to obtain the exponent $\nu$. From scaling relations,
once we dispose of $\eta$ and $\nu$ all other static exponents are determined.

\begin{figure}[th]
\begin{center}
\epsfig{file=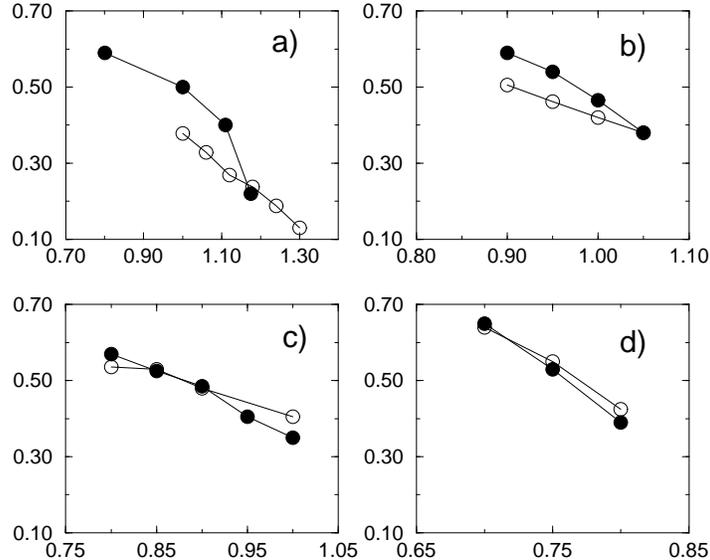,height=8cm,bbllx=18,bblly=18,bburx=552,bbury=462}
\end{center}
\caption{$\eta_1$ ($\circ$) and $\eta_2$ ($\bullet$) vs $T$ for various 
distributions. a) $\pm J$, b) Uniform c) Gaussian and d) decreasing 
exponential. Note that the scale on the $x$ axis is different for each plot.
Error bars on individual $\eta$ points are about $\pm$ 0.02.}
\label{fig:1}
\end{figure}

        We show in figure~\ref{fig:1}
estimates for $\eta_1(T)$ and $\eta_2(T)$ for the $3d$ $\pm J$ ISG 
calculated using data taken from the literature : $x(T)$ from \cite{3}, 
$h(T)$ from \cite{8,9}, and the spin glass susceptibilities for different
 assumed values of $T_g$; ($T_g=1.0$ from the data given in \cite{5}, 
$T_g = 1.11$ from \cite{6}, and $T_g=1.175$ from \cite{3}). There is a
well defined crossing point with $T_g=1.165 \pm 0.01$ and $\eta=-0.245 
\pm 0.02$. Using the curve for $z(T)$ from equation \ref{eq:8} we estimate 
$z=6.0 \pm 0.2$.

        The values obtained in this way are at least as precise as previous
estimates and are very close to the central  values given by Ogielski \cite{3}
($T_g=1.175 \pm 0.025$, $\eta=-0.22 \pm 0.05$, $z=6.0\pm0.8$), corroborating
 his
analysis. On the other hand the $T_g$ is marginally outside the error bars
quoted by Kawashima and Young ($T_g=1.11 \pm 0.04$) who use extensive Binder
cumulant data \cite{6}. The difficulty in applying this latter method to the
 $3d$ $\pm J$ ISG case is that the $g_L(T)$ curves lie very close together 
below $T_g$ so
the intersection point is sensitive to small changes in individual $g_L$
curves. Even with extreme statistical accuracy, small corrections to finite
size scaling (invoked as a possibility in \cite{6}) can change the apparent
position of the intersection point significantly. The results of ref \cite{6}
could be rendered consistent with the present analysis if the $g_L$ values for
the smallest samples studied were affected by corrections to finite size
scaling at the 1\% level.

        The present method is much less sensitive to problems of
systematics than are either of the other techniques outlined above. First,
both $x$ and $h$ are determined using "large" samples so finite size
corrections should be unimportant \cite{8,9}. Secondly
$h$ is measured out of equilibrium and so is not subject to the problems of
long equilibration times. The fact that no preparatory anneal is required
also means that the measurements are economical in computer time. The
measurements of $x$ need careful equilibration but systematic tests using
successsively longer preliminary anneals allow one to obtain reliable
values. Numerical data \cite{3, 10} show that in ISGs $q(t)$ already takes 
on the asymptotic form, equation \ref{eq:3}, from quite early times 
$t \simeq 2$ MCS (Monte Carlo Steps), and that sample to sample variations 
in the values of $x$ are small so extensive averaging over very large 
numbers of samples (an essential condition for good $g_L$ data) is 
unnecessary. Thus the curve $\eta_1(T)$ can be established accurately 
with moderate numerical effort and minimal systematic error. For the 
finite size scaling data from which $\eta_2(T)$ is deduced, thorough 
equilibration is necessary but by studying pairs of replicas \cite{5} 
and again testing with increasing anneal times it is  easier to obtain 
accurate values of $<q(t)^2>$ than the combination of moments which
constitute the Binder cumulant. Again, the sample to sample variability is
much less for $<q(t)^2>$ than for the Binder cumulant. In the $3d$ $\pm J$ 
ISG the two curves $\eta_1(T)$ and $\eta_2(T)$ intersect cleanly, 
figure~\ref{fig:1}, so the determination of the crossing point should not 
be very sensitive to minor deviations from scaling or small statistical 
uncertainties. Finally, no hypothesis is made concerning the way divergences 
occur except the essential assumption that standard scaling rules (as 
opposed to universality rules) hold. The excellent overall agreement between
Ogielski's estimates \cite{3}  and the present ones gives considerable
confidence in the general coherence of the standard scaling approach and
appear to make any exotic scaling assumption unnecessary.

        We therefore consider that both $\eta_i(T)$ curves can be 
calculated with little in the way of disguised systematic errors; as they 
stand the $T_g$ and exponent values that we quote should not only be precise 
but reliable. 

        We have made further simulations on another $3d$ ISG with $\pm J$
interactions; this is the the fully frustrated system with 20\% random bond
disorder that we studied in \cite{10}. We already established an accurate value
of $T_g$ ($T_g=0.96$) for this spin glass from Binder cumulant measurements, 
and we now have measured the exponents $x$ and $h$ at $T_g$ together with an 
estimate of $\eta$ from  the spin glass susceptibility (see 
Table~\ref{table:1}). The data are very consistent with each other and lead 
to an $\eta$ value which is less negative and a $z$ value which is smaller as 
compared with those of the standard $\pm J$ ISG. This difference already 
indicates the non-universality of these two exponents in $3d$ ISGs.

        We have also carried out extensive simulations on $3d$ ISG systems
with different sets of near neighbour interactions. For the $3d$ ISGs with
near neighbour Uniform, Gaussian and decreasing Exponential interactions
(see \cite{11} for the definitions of the distributions with the correct
normalizations), the data are shown in figure~\ref{fig:1}. Simulations
were done on samples with $L=16$ for $x$, $L=10$ for $h$, and samples 
from $L=2$ to $L=6$ for $<q(t)^2>$. Careful anneals were carried out where 
appropriate, checked by the prescription given in \cite{5}. At each 
temperature, 10 samples were used for $x$, 500 for $h$ and 2000 to 200 
depending on $L$ for $<q(t)^2>$. We estimate that the $\eta_1(T)$ curves 
are on large enough samples for there to be virtually no finite size 
correction, so the values can be taken as definitive (apart from statistical 
errors), but measurements on larger samples could modify the $\eta_2(T)$ 
curves marginally. It can be seen that the $\eta(T)$ curves again cross 
cleanly for the Uniform case with a more negative $\eta$
than for the $\pm J$ case. However for the Gaussian and Exponential cases it
turns out that the two curves are much more similar to each other making it
difficult to identify $T_g$ precisely; for these distributions we have to fall
back on an alternative method to estimate $T_g$.

        The Migdal-Kadanoff (MK) scaling approach is known to give
reasonable values of the ordering temperature for Ising spin glasses
\cite{12,13,14}. We have followed the particular method used by Curado and
Meunier \cite{14} but with improved statistical accuracy. It turns out that 
with a scale factor $b=2$ the MK estimate for the $3d$ $\pm J$ ISG $T_g$ 
is $1.16 \pm 0.01$, precisely the same as the value we have obtained above 
from the simulations. This perfect agreement is certainly fortuitous 
(though in $4d$ where the MK method should be much poorer, the disagreement 
in $T_g$ between the $b=2$ MK estimate and an accurate simulation value 
is only 15\% \cite{15}), but we argue that as agreement happens to be 
excellent for the $\pm J$ case, if we apply the same method with the same 
scale factor $b$ to other $3d$ ISGs with different sets of interactions, 
we should obtain $T_g$ estimates which should
again be very close to the real values. We obtain MK $T_g$ values which are
1.00, 0.88, and 0.72 for the Uniform, Gaussian and Exponential
distributions respectively \cite{15}. The Uniform distribution value is in good
agreement with the simulation value and the other two $T_g$ values are within
the range of $T$ where the simulation curves for $\eta_1(T)$ and $\eta_2(T)$ 
overlap. The Gaussian $T_g$ and $\eta$ are in good agreement with earlier 
estimates \cite{5}. Putting uncertainties at $\pm 0.05$ for possible 
systematic errors in the Gaussian and Exponential MK $T_g$ estimates, 
we obtain the set of exponent estimates shown in Table~\ref{table:1}.
\begin{table}[ht]
\begin{center}
 \begin{tabular}{||c|c|c|c|c|c||} 
 \hline
System  & $T_g$          & $x(T_g)$  & $h(T_g)$ &    $\eta$       &   $z$         \\ \hline \hline
$\pm J$  & $1.165\pm0.01$ & 0.064     &  0.38    & $-0.245\pm0.02$ & $6.0\pm0.2$   \\ \hline
FFd0.2  & $0.96\pm0.02$  & 0.091     &  0.437   & $-0.12\pm0.02 $ & $4.85\pm0.3$  \\ \hline
U       & $1.05\pm0.03$  & 0.054     &  0.41    & $-0.375\pm0.03$ & $5.8\pm0.5$   \\ \hline
G       & $0.88\pm0.05$  & 0.035     &  0.355   & $-0.50\pm0.04$  & $7.1\pm0.6$   \\ \hline
Exp     & $0.72\pm0.05$  & 0.02      &  0.275   & $-0.62\pm0.12$  & $9.5\pm0.7$   \\ \hline
  \end{tabular}
\end{center}
\caption{Temperature of transition and critical exponents for several 
distributions. The  distributions are in order ($i$) random $\pm J$ 
interactions, ($ii$) Fully Frustrated lattice with 20\% disorder 
\protect\cite{10}, ($iii$) random Uniformly distributed interactions, 
($iv$) random Gaussian interactions, ($v$) random Decreasing Exponential
interactions}
 \label{table:1}
\end{table}

                According to the usual universality rules, the form of the
interaction distribution should not be a pertinent parameter as concerns
the critical exponents. Here we find that $3d$ ISG systems which differ only
by this distribution function show quite different $\eta$ and $z$ values, 
Table~\ref{table:1}. The results indicates a breakdown of conventional 
universality in $3d$ ISGs.

\begin{figure}
\begin{center}
\epsfig{file=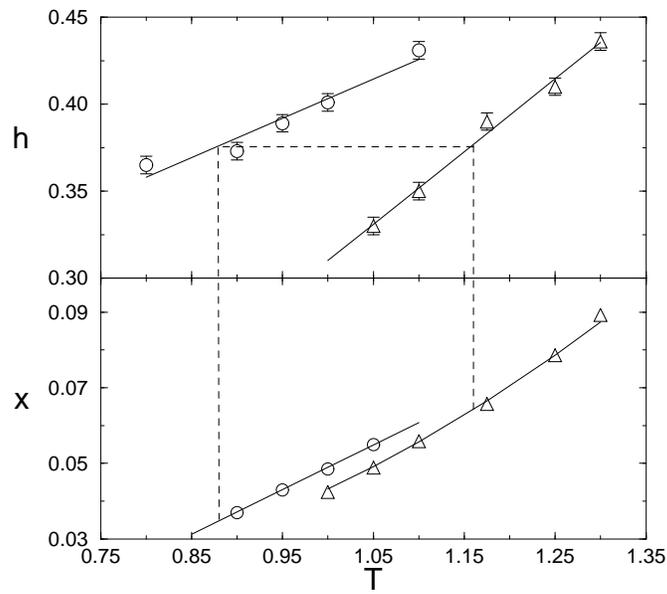,height=8cm,bbllx=18,bblly=18,bburx=552,bbury=462}
\end{center}
\caption{$h(T)$ and $x(T)$ for $\pm J$ ($\bigtriangleup$) and Uniform ($\circ$)
distributions. The temperature scale is common. The dashed line corresponds to
the example given in the text}
\label{fig:2}
\end{figure}

                In order to show that the apparent non-universality is not
an artefact, we will turn back to the raw $x$ and $h$ data for the $\pm J$
and Uniform cases. In figure~\ref{fig:2}
we have plotted the values of these parameters as a function of $T$; the error
bars are about $\pm 0.005$ for $h$ and $\pm 0.002$ for $x$. If universality 
holds
\begin{equation}
h(T_g(U)) \equiv h(T_g(J))
\label{eq:10}
\end{equation}
and
\begin{equation}
x(T_g(U)) \equiv x(T_g(J)).
\label{eq:11}
\end{equation}
By inspection, whatever trial value $T^*$ we choose for $T_g(J)$ within the 
generous limits $T^*=$ 1.0 to 1.3 provided by the figure, the 
relation~\ref{eq:10} leads to us to a $T^*_g(U)$ such that $x(T^*_g(U))$ is
considerably smaller than $x(T^*_g(J))$. For instance with $T_g^*(J)=1.16$,
$T_g^*(U)=0.88$, $x(T_g^*(J))=0.064$, $x(T_g^*(U))=0.036$. The data cannot 
satisfy \ref{eq:10} and \ref{eq:11} simultaneously, demonstrating 
non-universality.

        For the $2d$ regularly frustrated systems which show continuous
variation of critical exponents, the breakdown of universality is
necessarily associated with the existence of a marginal operator \cite{16} and
it has been pointed out that when breakdown occurs, it does so in Ising
systems having more than two ground states \cite{17} and hence with $n$, the
number of components of the order parameter, greater than 1 \cite{18}. On the
Parisi image of finite dimension ISGs \cite{19}, $n$ is essentially infinite; 
it would be of interest to identify possible marginal operators. We can note
that in the regularly frustrated $2d$ systems quoted above, $\nu$ varies
continuously but $\eta$ is constant so "weak universality" \cite{20} still 
holds. This is not the case for the randomly frustrated systems we have 
studied.

        It would appear that universality breakdown could be much more
prevalent than was suspected, and it may well be the rule rather than the
exception at spin glass or glass transitions.

        We would like to thank D$^r$ N. Kawashima for permission to quote
unpublished data. Simulations were carried out thanks to time allocations 
from IDRIS (Institut du D\'eveloppement des Ressources en Informatique 
Scientifique) and TRACS, University of Edinburgh. L.W.B. gratefully 
acknowledges support from TRACS.

\end{document}